# Dielectric Thickness Dependence of Carrier Mobility in Graphene with HfO$_2$ Top Dielectric

Babak Fallahazad[1], Seyoung Kim[1], Luigi Colombo[2], Emanuel Tutuc[1]

[1]*Microelectronics Research Center, The University of Texas, Austin, Texas, 78758, USA*
[2]*Texas Instruments, Inc. 12500 TI Boulevard Dallas, TX 75266, USA*

We investigate the carrier mobility in mono- and bi-layer graphene with a top HfO$_2$ dielectric, as a function of the HfO$_2$ film thickness and temperature. The results show that the carrier mobility decreases during the deposition of the first 2-4 nm of top dielectric and remains constant for thicker layers. The carrier mobility shows a relatively weak dependence on temperature indicating that phonon scattering does not play a dominant role in controlling the carrier mobility. The data strongly suggest that fixed charged impurities located in close proximity to the graphene are responsible for the mobility degradation.

Graphene, a mono-layer of carbon atoms arranged in a honeycomb lattice, has recently been the subject of considerable theoretical and experimental interest.[1-2] The potential for scalability to nanometer dimensions, high carrier mobility,[3] combined with chemical and mechanical stability[4] make graphene a promising candidate for nanoelectronic devices.[5] While the intrinsic carrier mobility in graphene is very high, with values of ~200,000 cm$^2$/Vs reported in suspended graphene,[6] scattering by charged impurities,[7] surface roughness,[8] and phonons[9] reduce the mobility in graphene devices integrated with dielectrics. High-$k$ dielectrics, such as HfO$_2$, are essential components in aggressively scaled complementary metal-oxide-semiconductor (CMOS) devices,[10] and will likely play a key role for graphene-based devices. Understanding the impact of the dielectric on mobility in graphene is not only technologically relevant, but can shed light on the scattering mechanism in this material.

A high-$k$ dielectric medium is expected to better screen charged impurities located in proximity to a graphene layer,[11] leading to higher mobilities. Several experimental studies have examined the impact of a top medium- or high-$k$ dielectric[12-15] on the carrier mobility in graphene. Jang *et al.*, deposited layers of ice ($k$≈3.2) on monolayer graphene at ~77K, and observed a gradual mobility increase (up to 30%) as a function of the ice thickness.[12] Chen *et al.* demonstrated a mobility enhancement at room temperature with deposition of high dielectric constant liquids ($k$≈32-189) on graphene devices fabricated on SiO$_2$/Si substrates.[13] Ponomarenko *et al.* observed a mobility enhancement factor of 2 and 1.5 when covering graphene with glycerol ($k$≈45) and ethanol ($k$≈25) respectively.[14] On the other hand, the carrier mobility in devices using conventional medium- or high-$k$ dielectrics, such as Al$_2$O$_3$ or HfO$_2$ are typically lower than the mobility of back-gated graphene devices. Indeed, the highest reported mobility values in graphene devices with Al$_2$O$_3$ ($k$≈6) top dielectric is ~8600 cm$^2$/Vs,[15] and typical mobility values for graphene with HfO$_2$ top dielectric are below 5,000 cm$^2$/Vs.[16]

Here we report the dependence of the graphene carrier mobility measured at room temperature on the thickness of a top HfO$_2$ dielectric grown by Atomic Layer Deposition (ALD). We observe a considerable mobility reduction of about 50% of the initial value after the first 2-4nm of metal oxide deposition. The temperature dependence of the mobility reveals a modest change down to 77K, suggesting that phonon scattering does not play a dominant role in our devices. The data indicates that fixed charged impurities located in close proximity of the graphene layer reduce the mobility. We speculate that these charges stem from charged point defects, such as oxygen vacancies, in the ALD high-$k$ dielectric.

Our device fabrication starts with 285 nm-thick thermally grown SiO$_2$ on an *n*-type Si (100) wafer, with an arsenic doping concentration of $N_D$>10$^{20}$ cm$^{-3}$. Graphene flakes are mechanically exfoliated from natural graphite crystals, and mono- and bi-layer graphene pieces are identified and isolated on SiO$_2$, using Raman spectroscopy[17] and optical contrast.[18] Four point and Hall bar device geometries are then defined using electron beam lithography (EBL) and oxygen plasma etching. A second EBL step, 50 nm-thick Ni deposition, and lift-off are used to define the metal contacts.

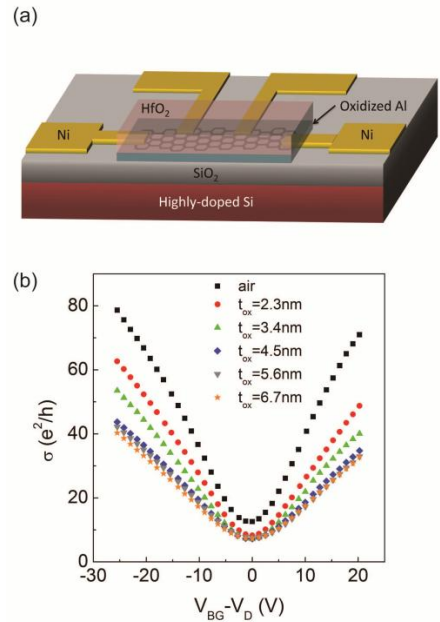

Fig. 1. (a) (color online) Schematic of a graphene device on a 285 nm thick SiO$_2$ bottom dielectric, and with a top dielectric stack consisting of an Al buffer layer, followed by ALD of HfO$_2$. (b) $\sigma$ vs. $V_{BG}$, determined using four-point measurements, for different top dielectric stack thicknesses ($t_{ox}$). The *x*-axis is offset by the Dirac voltage ($V_D$), at which the graphene conductivity is minimum.

Prior to the HfO$_2$ deposition, a thin (~1.5nm) pure Al film (99.999%) is deposited by e-beam evaporation to provide nucleation sites for the ALD process.[10,15] The Al layer becomes oxidized once the deposition chamber is vented and exposed to the air,[19] and forms a thin metal-oxide interface film. The sample is then transferred to the ALD chamber for successive, ~1nm thick HfO$_2$ deposition cycles. The HfO$_2$ ALD was performed at a temperature of 200°C using Tetrakis[EthylMethylAmino]Hafnium (TEMAH) and H$_2$O as precursors, without any post-deposition annealing. Figure 1(a) shows the schematic of a back-gated graphene device with a top dielectric. The stack relative dielectric constant ($k$~16) is measured by adding a top metal gate and comparing the relative capacitance of the top and bottom gates.[15]

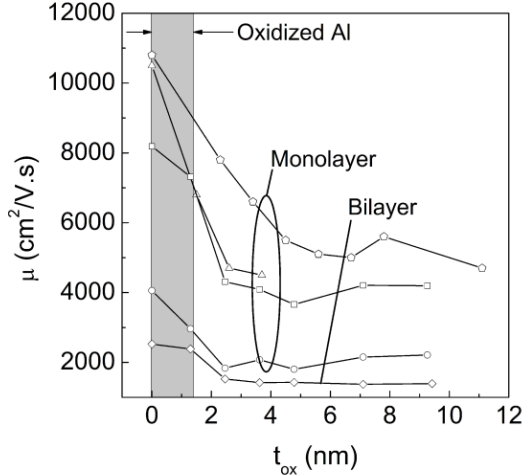

Fig. 2. $\mu$ vs. $t_{ox}$ for four mono-layer and one bi-layer graphene devices. The mobility decreases steeply after the first 2-4 nm of oxide deposition, and remains constant for thicker dielectric films.

The four-point device conductivity ($\sigma$) measured as a function of the back-gate bias ($V_{BG}$), at room temperature, under vacuum, and for different total top dielectric thicknesses ($t_{ox}$) are shown in Figure 1(b). The data are measured before the dielectric deposition and also after each incremental HfO$_2$ deposition. The measurement shows the minimum conductivity at the charge neutrality (Dirac) point drops from $12e^2/h$ to $8e^2/h$ after the Al deposition and the first HfO$_2$ layer, and is unchanged with further HfO$_2$ depositions. Using the $\sigma$ vs. $V_{BG}$ data, the carrier mobility ($\mu$) is extracted after each HfO$_2$ deposition which in turn provides us with the dielectric stack thickness ($t_{ox}$) dependence of the carrier mobility. The carrier mobility is calculated from the linear slope of $\sigma$ vs. $V_{BG}$ data, using $d\sigma/dV_{BG} = \mu \cdot C_{ox}$; where $C_{ox} \approx 12$nF/cm$^2$ is the SiO$_2$ bottom dielectric capacitance. To avoid the non-linearity around the minimum conductivity point when extracting $d\sigma/dV_{BG}$, we exclude a 6V voltage window centered at the charge neutrality back-gate bias ($V_D$), and average out the slopes of $\sigma$ vs. $V_{BG}$ over a 25V $V_{BG}$ window, on the electron and hole branches; the electron and hole mobilities differ by less than 10%. We note that our approach of using the slope of $\sigma$ vs. $V_{BG}$ data to extract the mobility neglects short-range scattering, e.g. from neutral impurities.[20] Taking into account the effect of short-range scattering, the extracted mobility values in our samples would change only slightly, by ~10%.

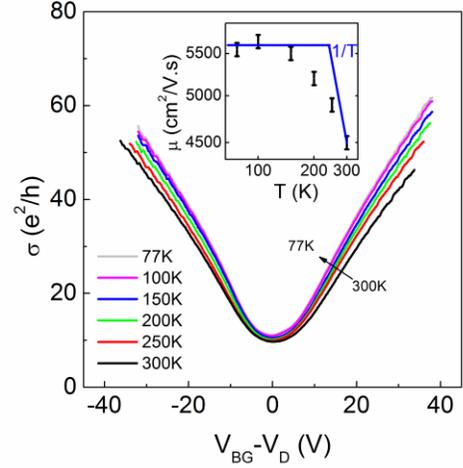

Fig. 3. (color online) $\sigma$ vs. $V_{BG}$ measured at different temperatures ($T$) for a graphene device with an 11nm HfO$_2$ top dielectric. The $V_{BG}$ values are offset by the charge neutrality voltage ($V_D$). The arrow indicates the evolution of $\sigma$ vs. $V_{BG}$ data sets with decreasing $T$. Inset: $\mu$ vs. $T$ for the same device. The relatively weak $T$-dependence suggests that phonons are not the mobility limiting factor in these devices.

The data in Figure 2 shows the mobility ($\mu$) vs. dielectric stack thickness ($t_{ox}$), measured for four mono-layer and one bi-layer devices at room temperature. A mobility drop is observed after the formation of the oxidized Al buffer layer, and also the deposition of the first 1-2 nm of HfO$_2$. To further investigate the scattering mechanism in graphene with HfO$_2$ top dielectric, in Fig 3 we show $\sigma$ vs. $V_{BG}$ for a mono-layer graphene with an 11 nm-thick HfO$_2$ top dielectric film, at different temperatures ($T$). The $\mu$ vs. $T$ data shown in Fig. 3 shows a $\mu \propto 1/T$ dependence at higher $T$ values, consistent with acoustic phonon scattering,[9] followed by a saturation at the lowest $T$. These data reveal a weak temperature dependence which indicates that phonon scattering is not dominant in our devices. Since the surface roughness is not expected to change with the top dielectric deposition, the $\mu$ vs. $t_{ox}$ data of Fig. 2 combined with the $\mu$ vs. $T$ of Fig. 3 strongly suggest that fixed charged impurities located in the high-$k$ dielectric, and in close proximity to the graphene layer are responsible for the mobility degradation. Next we address the origin of these additional charged impurities that accompany the top dielectric deposition. The metal-oxide dielectrics, either Al$_2$O$_3$ or HfO$_2$, are deposited at room temperature or 200°C respectively. Dielectrics deposited at low temperatures, such as the ALD process used here, are generally not stoichiometric, but oxygen deficient. We speculate that these charged impurities are point defects, such as charged oxygen vacancies.[21-22] Indeed, the oxygen vacancies form donor levels closer in energy to the HfO$_2$ conduction band, and higher than the graphene Fermi level. Similar to a metal–high-$k$ dielectric stack,[21-22] the electrons tunnel out of the dielectric and into the graphene in order to bring in equilibrium the Fermi levels in graphene and HfO$_2$ [Fig. 4(a)], and the point defects in close proximity to the graphene layer become charged, which in turn reduces the carrier mobility.

To quantify the above argument, we employ the Boltzmann transport formalism where charged impurity screening is treated within the random phase approximation.[11] We use $\mu$ vs. $t_{ox}$ data of Fig. 2 to estimate the charged impurity areal density ($n_{imp}$) from $n_{imp} = \frac{2e}{h}\frac{1}{\mu}\frac{1}{F_l(\alpha)}$, where $e$ is the electron charge, $h$ is Plank's constant, $\alpha$ is the dimensionless coupling constant $\alpha = \frac{2e^2}{(\kappa_1+\kappa_2)\hbar v_f}$, $v_f = 1.1\times 10^6$ m/s is the graphene Fermi velocity, $k_1=16$ and $k_2=3.9$ are the dielectric constants of top and bottom oxides, and

$$F_l(\alpha) = \pi\alpha^2 + 24\alpha^3(1-\pi\alpha) + \frac{16\alpha^3(6\alpha^2-1)\arccos(1/2\alpha)}{\sqrt{4\alpha^2-1}}.$$

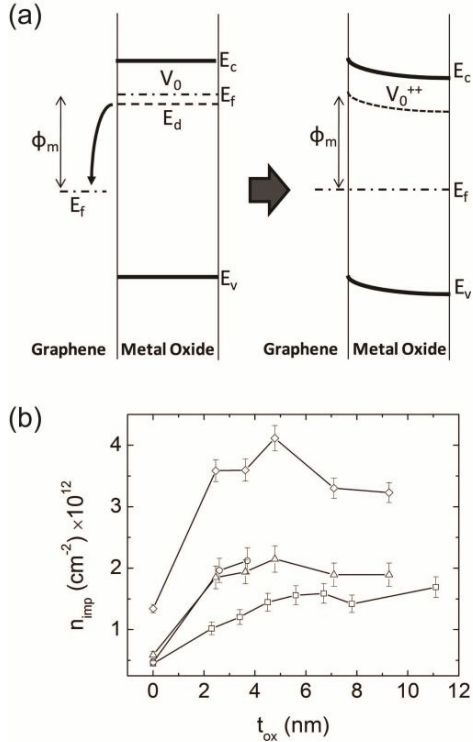

Fig. 4. (a) Schematic of band diagram for a stack consisting of metal-oxide deposited on graphene. The oxygen vacancies, inherent for dielectrics deposited at low temperatures become ionized in the proximity of the interface, creating fixed charged impurities which reduce the mobility (Reproduced from Ref. [22]). (b) $n_{imp}$ vs. $t_{ox}$ data for four graphene monolayers.

Figure 4 shows $n_{imp}$ vs. $t_{ox}$ for four monolayer graphene devices. These data suggest that the dielectric deposition increases the charged impurity concentration by ~1.5-$4\times10^{12}$ cm$^{-2}$. These values are in good agreement with previous studies which examined the thermochemistry of metal-oxide-semiconductor structures using $HfO_2$ on Si.[21-22] In summary, we studied the mono- and bi-layer graphene mobility dependence on the thickness of a top high-$k$ metal-oxide dielectric. Four-point, gate-dependent measurements show that graphene mobility decreases after 2-4nm metal-oxide dielectric deposition and remains constant if the dielectric thickness is further increased. The mobility temperature dependence suggests that phonons are not the dominant scattering mechanism in these devices, indicating that additional charged impurities located in close proximity to the graphene layer are introduced during dielectric deposition. We speculate that positively charged oxygen vacancies, ubiquitous in high-$k$ dielectrics, are the mobility limiting factor in our devices.

We thank S. Guha for discussions. This work is supported by NRI-SWAN, and by DARPA Contract FA8650-08-C-7838 through the CERA program, and IBM-UT subcontract agreement W0853811.